# Understanding Binary Code Similarity for Real-World Vulnerability Detection: A Large-Scale Empirical Study

JINGDONG GUO*, Institute of Information Engineering, CAS, China and University of Chinese Academy of Sciences, China
CHAOPENG DONG*, Hangzhou Dianzi University, China, Institute of Information Engineering, CAS, China, and University of Chinese Academy of Sciences, China
YIMO REN, Institute of Information Engineering, CAS, China and University of Chinese Academy of Sciences, China
SIYUAN LI, Shandong University, China
JIE LIU, HONG LI†, and HONGSONG ZHU, Institute of Information Engineering, CAS, China and University of Chinese Academy of Sciences, China

Firmware lies at the heart of IoT devices. Its development depends heavily on third-party libraries (TPLs), which greatly accelerate the process but simultaneously introduce associated vulnerabilities. Binary Code Similarity Detection (BCSD) is an effective technique for identifying vulnerabilities in firmware by comparing pairs of code segments. However, existing studies either evaluate their performance only on small-scale datasets or lack diversity in terms of vulnerabilities, TPLs, and firmware. Consequently, a comprehensive understanding of BCSD for real-world vulnerability detection remains absent.

To bridge this gap, we conduct a large-scale study of vulnerability detection across 60,000 firmware images from 200 vendors using BCSD. Rather than introducing a novel model, we examine the influence of four key factors—vulnerable function versions, vulnerability search space, function sizes, and compilation toolchains on BCSD performance. Our results reveal that these factors substantially affect performance, often by wide margins. To address this, we propose a build-aware query strategy that derives queries from representative real-world binaries, effectively closing the gap and raising the mean reciprocal rank (MRR) from 0.818 to 0.981. Furthermore, we demonstrate that a TPL-aware, two-stage search process significantly enhances accuracy, improving MRR by 18.5% by limiting the search space.

*Both authors contributed equally to this research.
†Corresponding author.

Authors' Contact Information: Jingdong Guo, guojingdong@iie.ac.cn, Institute of Information Engineering, CAS, Beijing, China, School of Cyber Security and University of Chinese Academy of Sciences, Beijing, China; Chaopeng Dong, dongchaopeng@hdu.edu.cn, School of Cyberspace, Hangzhou Dianzi University, Hangzhou, Zhejiang, China and Institute of Information Engineering, CAS, Beijing, China, School of Cyber Security and University of Chinese Academy of Sciences, Beijing, China; Yimo Ren, renyimo@iie.ac.cn, Institute of Information Engineering, CAS, Beijing, China, School of Cyber Security and University of Chinese Academy of Sciences, Beijing, China; Siyuan Li, lisiyuan201@mails.ucas.ac.cn, School of Cyber Science and Technology, Shandong University, Jinan, Shandong, China; Jie Liu, liujie1@iie.ac.cn; Hong Li, lihong@iie.ac.cn; Hongsong Zhu, zhuhongsong@iie.ac.cn, Institute of Information Engineering, CAS, Beijing, China, School of Cyber Security and University of Chinese Academy of Sciences, Beijing, China.

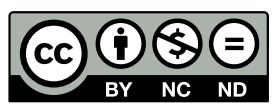

## 1 Introduction

The Internet of Things (IoT) has become deeply embedded in modern society, bringing significant convenience to our lives, such as smart homes, critical infrastructure, and industrial control systems. According to the report of Statista [41], the number of connected IoT devices will exceed 30 billion by 2030. Nevertheless, the extensive use of IoT devices also poses security threats [1]. For example, the Mozi botnet [21] is a P2P IoT botnet that exploits weak credentials and vulnerabilities in routers and DVRs. Its DHT-based structure makes it resilient and enables large-scale DDoS attacks.

Firmware constitutes the foundation of IoT devices, orchestrating both hardware initialization and essential system services [22]. In the present era, to accelerate development and satisfy increasingly diverse feature demands, IoT developers frequently incorporate third-party libraries (TPLs) into firmware, such as OpenSSL for encryption [36] and Zlib [53] for compression. Yet this integration is double-edged: while it expedites development, it simultaneously introduces TPL-related vulnerabilities, thereby exposing IoT devices to potential exploitation by attackers. Even more troubling, a vulnerability in a widely adopted TPL (e.g., OpenSSL) can silently proliferate across millions of devices from diverse vendors, owing to the pervasive reuse of identical code [51].

Binary Code Similarity Detection (BCSD) [17, 39] is a precise and efficient technique employed to identify vulnerabilities arising from code reuse. Given a known vulnerable function from a TPL, BCSD compares it against functions within the firmware, measuring their similarity to identify vulnerable functions. With the development of deep learning and natural language processing (NLP) techniques, BCSD are gradually shifting from traditional rule-based [7, 9] to learning-based [18, 37, 47–49], which shows greater scalability and accuracy. Figure 1 illustrates the pipeline of vulnerability detection using learning-based BCSD methods, which can be broadly divided into three principal stages:

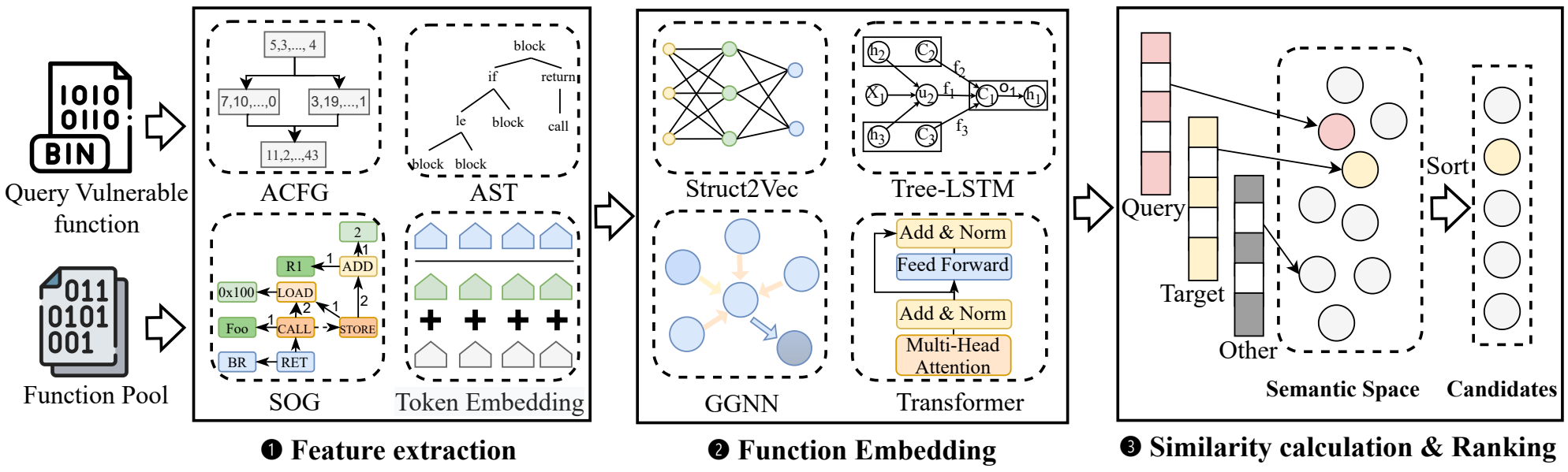

Fig. 1. The BCSD pipeline for vulnerability discovery. A vulnerable function (query) is input to the BCSD engine, which searches a corpus of target binaries and outputs a ranked list of potential matches.

- ❶ **Feature Extraction.** Given a query (known) vulnerable function and a pool composed of a set of functions, the features are extracted and represented in various formats (e.g., ACFG).
- ❷ **Function Embedding.** The extracted features are input into a neural network (e.g., Struct2Vec), which produces function embeddings that encapsulate function semantics in the form of high-dimensional vectors.
- ❸ **Similarity Calculation and Ranking.** Within the semantic space, the query function embedding is compared against other embeddings by calculating cosine similarities, after which the candidates are ranked according to their similarity scores.

Although a number of studies [9, 11, 14, 18, 24, 31, 37, 47–49] have applied BCSD to vulnerability detection, investigations into its effectiveness on real-world firmware remain limited for several reasons. *(1) Most existing works conduct real-world analysis only on a small set of firmware images or third-party libraries (TPLs) [9, 18, 24, 31, 37, 49], often restricted to a narrow range of device types or vendors.* As a result, the evaluation lacks diversity in both firmware and TPL ecosystems, making it difficult to generalize the conclusions to the heterogeneous landscape of real-world IoT and embedded systems. *(2) These studies typically assess BCSD using only a small number of vulnerabilities (e.g., 10–20) [11, 14, 47, 48].* Such a limited scope provides only partial evidence of effectiveness, leaving open questions about scalability, robustness, and applicability to large-scale vulnerability detection scenarios where various types of vulnerabilities may exist. In addition, existing real-world analyses are typically conducted to demonstrate the effectiveness of proposed approaches. Yet, they often lack a thorough examination of the underlying factors that may influence BCSD performance. Specifically, do variations across different versions of TPLs influence detection outcomes? Does the static linking of TPLs alter the results? How does the disparity in function size impact the performance of BCSD methods? Beyond factors extensively examined in prior research, such as compilers, optimization levels, and architectures, are there still any other factors that exert a significant influence on the efficacy of BCSD approaches? Therefore, a comprehensive understanding of how BCSD addresses vulnerability detection in real-world applications is still lacking.

To bridge this gap and conduct a large-scale systematic study on BCSD for real-world vulnerability detection, we have two key challenges as follows:

- **C1 The Construction of Firmware Dataset.** The firmware dataset forms the cornerstone of our empirical study; however, obtaining it is far from straightforward due to several practical challenges. First, firmware images are distributed across a highly diverse set of sources, including official vendor websites, ftp sites, and third-party repositories, each of which adopts its own release practices, formats, and versioning policies. This heterogeneity complicates the process of building a unified and representative dataset. Second, many vendors employ sophisticated anti-scraping mechanisms [43] such as rate limiting, dynamic content generation, or access restrictions, which hinder automated collection at scale.
- **C2 The Construction of Vulnerability Dataset.** Each vulnerability in our dataset consists of three primary components: the vulnerable function name, the affected versions, and the corresponding binary function used for matching. The first two components are generally described within CVE entries, which are expressed in unstructured natural language and thus challenging to extract accurately. The third component, however, demands not only the identification and retrieval of the correct source versions but also the automated compilation of software across a broad spectrum of projects, compiler toolchains, optimization levels, and target architectures. In particular, build dependencies may be absent, or the target version may be so outdated that it necessitates an entirely different compilation environment.

In this paper, we seek to offer a comprehensive understanding of BCSD's effectiveness in real-world firmware vulnerability detection, while also investigating the key factors influencing its performance that have been overlooked in prior works [18, 44, 47, 48]. Our analysis method is as follows. **First**, to address the challenge of firmware dataset construction, we design a pipeline that automatically collects firmware images from diverse sources (e.g., official vendor websites and third-party repositories), employing customized strategies to bypass site-specific restrictions, and subsequently preprocesses the firmware to extract binaries for vulnerability detection. **Second**, to address the challenge of vulnerability dataset construction, we begin by collecting vulnerability data from NVD [35], and then extract the necessary information through a three-step process:

rule-based extraction, LLM-assisted extraction, and human verification. To obtain the corresponding vulnerable binary functions, we clone the source code of the projects and compile the affected versions for each vulnerability using automated compilation scripts supplemented by manual adjustments when required. **Third**, since our goal is to analyze the performance of state-of-the-art BCSD approaches in vulnerability detection, rather than to propose a new method that surpasses existing ones, we begin by examining BCSD techniques from the past five years that support cross-compilation detection. Afterward, we retrain and evaluate these approaches on the public binary dataset BinKit [24] for pairwise comparison. The most effective approach is subsequently selected to perform large-scale vulnerability detection, thereby minimizing unnecessary computation and saving time.

To provide a more comprehensive analysis, we examine the primary factors influencing the performance of BCSD approaches from four perspectives: 1) the impact of different versions of TPLs on BCSD; 2) the effect of nested TPLs on BCSD; 3) the influence of function size on BCSD; 4) the impact of compilation macros on BCSD; Based on the analysis, we reveals that using a version-matched query improves the Mean Reciprocal Rank (MRR) by 14% compared to the common practice of using the latest vulnerable function. Furthermore, by combining TPL detection with BCSD to narrow the search space, we improved the MRR by 5.4% over BCSD alone. We also quantified the variation in MRR across different target function sizes and identified "complexity valleys."Finally, our proposed build-aware query strategy (obtaining vulnerable function binaries from representative in-the-wild binaries by clustering compilation macros) proves to be very effective, improving the MRR from 0.818 to 0.981, an improvement of nearly 20%.

We summarize our main contributions as follows.

- We construct so far the largest firmware and vulnerability dataset, which includes 46,802 firmware images from 200 vendors, covering nine categories such as NAS, Camera, and Switch. From these firmware files, we extracted 4,411,056 binary files with a total of 547,578 unique hash values. Additionally, we compiled a vulnerable binary dataset containing 1,069 binaries with nine different TPLs, linked to 402 CVEs and 53 CWEs (e.g., memory leaks, code injection, and buffer overflows).To facilitate future research, we will open-source this dataset.
- We conduct the first large-scale analysis of BCSD applied to vulnerability detection in firmware. In this study, we investigate four critical factors that influence detection performance: TPL versions, static linking, function sizes, and macro definitions. These factors, largely overlooked in prior research, also play a pivotal role in affecting the effectiveness of BCSD for vulnerability detection.
- We propose a build-aware methodology to pinpoint vulnerable binaries, based on the clustering of compilation macros from representative vendors. This approach has proven highly effective and demonstrates robust generalization across diverse firmware.

## 2 Study Design

Figure 2 illustrates the overall architecture of our study, which consists of two primary stages: ❶ **Data Collection and Preprocessing** and ❷ **Vulnerability Detection and Evaluation**.

The stage ❶ is dedicated to gathering and preprocessing essential resources, including firmware images from multiple distribution channels, TPL source code from GitHub or official vendor websites, and vulnerability information from NVD [35]. Subsequently, we preprocess the collected data by extracting and filtering binaries from the firmware, compiling the vulnerable versions of TPL source code into binaries, and identifying the vulnerable functions by referencing their names, thereby constructing the vulnerability dataset.

Stage ❷ is designed to both detect vulnerabilities and evaluate the results from multiple perspectives. Using the binaries obtained in the previous stage and the optimal BCSD approach selected through the public binary dataset (i.e., BinKit), we construct the vulnerability dataset and function pool. With these, we perform vulnerability detection and address the four research questions, assessing the impact of TPL versions, TPL nesting, function size, and macros.

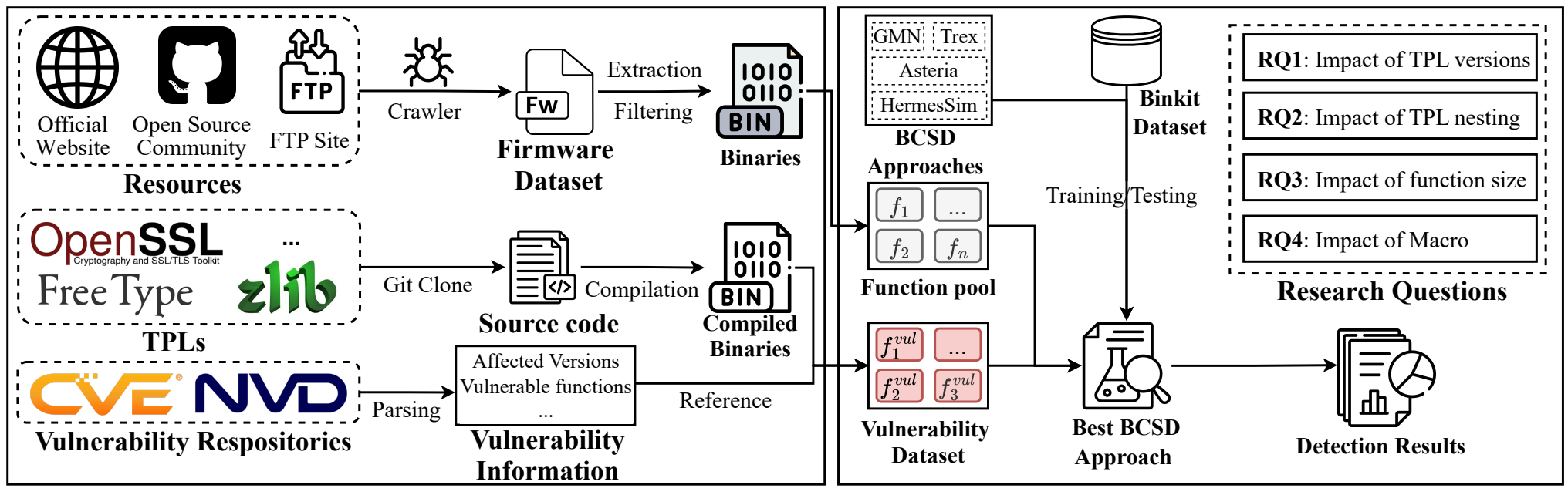

Fig. 2. Overview of our study architecture.

## 2.1 Data Collection and Preprocessing

To address the limitations of prior benchmarks, we constructed a large-scale, diverse, and realistic dataset derived entirely from real-world firmware.

*2.1.1 Firmware Dataset Construction.* We constructed our firmware *dataset* according to three principles that address C1:

- **Source discovery and normalization.**We collect firmware from official vendor portals, FTP sites, and vetted community mirrors. Per-vendor fetchers normalize metadata (vendor, product, version, URL, optional checksum) into a unified manifest, keeping the *dataset* consistent and searchable.
- **Responsible collection.**As vendors deploy rate limits, dynamic pages, and gated downloads, we (1) throttle with exponential backoff and vendor-specific caps; (2) render pages via a headless browser to capture official links; (3) use public logins/tokens where available; and (4) hand-queue CAPTCHA-guarded downloads. We do not bypass protections; items requiring extra credentials are marked ineligible.
- **Integrity checks and deduplication.**For each image we verify published checksums (when present) and compute a SHA-256 hash. We remove duplicates by (vendor, model, version, hash) and collapse repacks by comparing inner filesystem hashes, yielding a clean, non-redundant *dataset*.

**Preprocessing and extraction.**We first filter out non-firmware files, then unpack all images with Binwalk [19]. During extraction, we record the detected CPU architectures and filesystems for subsequent characterization. Encrypted or unknown filesystems are flagged and excluded. After unpacking, we retain executables and shared objects, discard non-binaries, and deduplicate binaries by content hash to obtain the analysis-ready *dataset*.

**Scale and characteristics.** From nearly 200 vendors (e.g., Hikvision, TP-LINK, Cisco) we collect **59,374** firmware images. We successfully unpack **46,802** images (yielding about 4.4M binary files after extraction). After filtering and deduplication, the final *dataset* contains **547,578** unique binaries. Architectures are dominated by **ARM (45.62%)** and **MIPS (44.91%)**, and **96.51%** of binaries are stripped, making the *dataset* both realistic and challenging.Detailed statistics are shown in Table 1

Table 1. Top 30 Vendors by Firmware Count and Other Statistics

| Vendor | #Firm | Category | | | Filesystem | | | | Architecture | | | | 32-bits (%) | Stripped (%) |
|---|---|---|---|---|---|---|---|---|---|---|---|---|---|---|
| | | NAS | Router | Other | Ext | jffs2 | Squashfs | Other | ARM | MIPS | x86 | Other | | |
| NETGEAR | 4,604 | 389 | 1,012 | 3,203 | 29 | 216 | 1,972 | 2,387 | 349,695 | 193,010 | 115,409 | 5,770 | 84.00% | 97.84% |
| D-Link | 3,677 | 20 | 265 | 3,392 | 117 | 20 | 922 | 2,618 | 69,801 | 87,441 | 1,894 | 27,031 | 98.32% | 98.98% |
| ASUS | 3,657 | 6 | 1,192 | 2,459 | 3 | 99 | 1,973 | 1,582 | 469,426 | 202,064 | 25 | 365 | 99.55% | 93.08% |
| Tomato | 3,319 | - | 18 | 3,301 | 0 | - | 3,316 | 3 | 30,469 | 546,131 | 0 | 0 | 100.00% | 100.00% |
| MikroTik | 2,894 | - | 2,774 | 120 | 4 | - | 2,803 | 87 | 55,318 | 208,507 | 58,860 | 123,426 | 99.76% | 97.62% |
| ZyXEL | 2,471 | 26 | 46 | 2,399 | 69 | 178 | 206 | 2,018 | 34,360 | 54,246 | 96 | 4,814 | 99.88% | 99.47% |
| Intel | 2,398 | - | 0 | 2,398 | 1 | 101 | 889 | 1,407 | 376,400 | 1 | 1,807 | 177 | 99.62% | 85.50% |
| Ubiquiti | 1,207 | - | 204 | 1,003 | 0 | 45 | 293 | 869 | 37,742 | 188,282 | 0 | 735 | 98.57% | 97.78% |
| Gargoyle | 1,046 | - | 12 | 1,034 | 9 | 15 | 907 | 115 | 3,942 | 156,278 | 2,167 | 0 | 99.70% | 99.85% |
| digi | 1,007 | - | 0 | 1,007 | 4 | 64 | 18 | 921 | 666 | 10,046 | 738 | 5,900 | 93.48% | 99.79% |
| TP-LINK | 934 | - | 583 | 351 | - | 10 | 528 | 396 | 17,179 | 37,756 | 118 | 368 | 100.00% | 98.97% |
| CANON | 812 | - | - | 812 | - | 0 | 0 | 812 | 148 | - | 10 | 0 | 93.67% | 94.30% |
| pelco | 800 | - | - | 800 | - | 122 | 475 | 203 | 222,969 | - | 3,307 | 165 | 96.98% | 96.35% |
| Schneider | 555 | - | - | 555 | - | 4 | 1 | 550 | 1,918 | 1 | 118 | 1,647 | 98.51% | 98.83% |
| TRENDnet | 543 | 2 | 131 | 410 | 15 | 15 | 235 | 278 | 13,250 | 18,899 | 350 | 34 | 100.00% | 98.96% |
| Supermicro | 528 | 0 | - | 528 | 0 | 0 | 0 | 528 | 7,667 | - | 192 | 24 | 99.53% | 100.00% |
| Dell | 526 | 1 | - | 525 | 1 | 6 | 9 | 510 | 4,695 | - | 3,766 | 4,490 | 72.85% | 99.95% |
| Conceptronic | 367 | 55 | 31 | 281 | 24 | 1 | 71 | 271 | 1,485 | 5,357 | 39 | 469 | 100.00% | 96.54% |
| Tplink | 366 | 0 | 0 | 366 | 3 | 11 | 133 | 219 | 3,279 | 5,913 | 20 | 2 | 100.00% | 99.91% |
| AirLive | 356 | 3 | 9 | 344 | 76 | 2 | 97 | 181 | 20,890 | 4,986 | 3,215 | - | 99.99% | 95.38% |
| Tenda | 342 | 0 | 76 | 266 | 0 | 3 | 154 | 185 | 10,949 | 10,234 | 0 | - | 100.00% | 100.00% |
| QNAP | 341 | 322 | - | 19 | 1 | - | 2 | 338 | 5,466 | 112 | 746 | - | 100.00% | 99.92% |
| UTT | 327 | - | - | 327 | 0 | - | 3 | 324 | 0 | 20,677 | 0 | 4,040 | 100.00% | 100.00% |
| Samsung | 319 | - | - | 319 | 8 | 5 | 20 | 286 | 38,102 | - | 104 | - | 99.99% | 99.73% |
| Hikvision | 317 | - | - | 317 | 49 | 3 | 0 | 265 | 4,631 | - | 0 | - | 100.00% | 100.00% |
| Wayos | 307 | - | - | 307 | 6 | 0 | 2 | 299 | 0 | 27,351 | 585 | 6 | 96.81% | 100.00% |
| idis | 290 | - | - | 290 | - | 257 | 1 | 32 | 861 | - | - | - | 62.83% | 100.00% |
| Sony | 277 | - | - | 277 | - | - | 4 | 273 | 905 | - | - | - | 100.00% | 100.00% |
| Pioneer | 272 | - | - | 272 | 11 | - | 9 | 252 | 6,678 | 9 | - | - | 100.00% | 96.83% |
| Belkin | 260 | - | 153 | 107 | 1 | - | 123 | 136 | 3,200 | 14,502 | 66 | - | 100.00% | 99.14% |
| Other | 6,124 | 353 | 580 | 5,191 | 155 | 197 | 1,472 | 4,300 | 217,565 | 183,011 | 36,453 | 8,238 | 89.34% | 98.33% |

*2.1.2 Vulnerability Dataset Construction.* We constructed our vulnerability *dataset* through a two-stage pipeline that addresses C2: (i) turning largely unstructured CVE texts into reliable tuples (vulnerable function names and affected version ranges), and (ii) compiling ground-truth ARM binaries despite heterogeneous and often outdated build environments.

*TPL Selection.* Building on prior large-scale analyses of TPL reuse in IoT firmware [3, 4], we select a core set that prior literature repeatedly reports as prevalent and security-critical: **OpenSSL** (TLS/SSL and Crypto), **BusyBox**[2] (system utilities), and **zlib** (compression). For practicality and to better cover real firmware attack surfaces, we additionally include six widely shipped TPLs spanning networking and parsers: **curl**[6] (network transfers), **libexpat**[28] (XML), **libpng**[29] and **libtiff**[30] (Image Codec), **FreeType**[13] (font rendering), and the **binutils** [15] suite (toolchain utilities). This literature-grounded selection reflects what prior work documents as common in the firmware, while ensuring our evaluation spans crypto, utilities, networking, and file parsers that frequently accumulate N-day CVEs.

*CVE Mining and Structuring.* For the nine TPLs, we curated **402** CVEs (2004–2024) from authoritative sources NVD [35]. Each CVE was normalized into two fields: *(a)* vulnerable function name(s),

*(b)* affected version range. Because the texts are unstructured, we use a lightweight, multi-step process focused on accuracy:

(1) *Regex pre-extraction.* Rule-based recognizers capture version ranges (e.g.,"1.1.0–1.1.1c", "≤ 1.0.2g") and function-like tokens as candidates.
(2) *Prompted LLM extraction.* A prompt-driven LLM converts CVE/advisory text into a structured record (functions and version bounds), with source spans for traceability.
(3) *Patch-message confirmation.* When an upstream fixing commit exists, we use the commit message to confirm the vulnerable function name(s).
(4) *Human verification.* Ambiguous items (e.g., aliased symbols) are manually reviewed; clear cases are sampled; all conflicts are 100% reviewed.

This procedure couples automation (regex+LLM) with commit-level confirmation and human checks, yielding high-confidence tuples summarized in **Table 2**.

*Compilation to ARM Binaries.* To obtain the *binary function* artifacts *used as the source of vulnerable binary functions for matching*, we compile representative snapshots for the **ARM** architecture—namely the oldest affected version, the latest affected version, and several intermediates. Compilation challenges (missing dependencies, legacy build tools, and environment drift) are handled as follows:

- **Containerized environments.** By default, we build in Docker on **Ubuntu 18.04** with **GCC 7.5.0** targeting `arm-linux-gnueabihf` and *default* compilation settings.
- **Legacy releases.** For older versions that cannot be built successfully in this setup, we switch to older Docker images (e.g., **Ubuntu 16.04** or **Ubuntu 14.04**) to match historical toolchains and system libraries, in order to address compilation-time challenges.
- **Missing dependencies.** When builds fail due to absent packages, we *manually compile and install the required dependencies* inside the container, then re-run the build with default settings.

Table 2. Summary of Vulnerability Metrics for TPLs. Abbreviations: "Vuln. Funcs" = Vulnerable Functions, "Comp. Vers." = Compiled Versions, "Comp. Bins" = Compiled Binaries.

| TPL | TPL Category | # CVEs | # CWEs | # Vuln. Funcs | # Comp. Vers. | # Comp. Bins |
|---|---|---|---|---|---|---|
| OpenSSL | TLS/SSL and Crypto | 118 | 32 | 139 | 141 | 282 |
| libtiff | Image Codec | 73 | 13 | 77 | 15 | 339 |
| FreeType | Font Rendering | 66 | 12 | 68 | 30 | 30 |
| curl | Networking/Data Transfer | 54 | 26 | 55 | 53 | 53 |
| libexpat | XML Parser | 28 | 11 | 32 | 16 | 16 |
| BusyBox | System Utilities | 20 | 11 | 20 | 14 | 14 |
| libpng | Image Codec | 19 | 9 | 26 | 23 | 23 |
| binutils | Binary Toolchain | 16 | 7 | 18 | 6 | 101 |
| zlib | Compression | 8 | 4 | 8 | 51 | 51 |

This strategy yields a comprehensive ARM binary set for the 402 CVEs across nine TPLs, with default build settings preserved and practical obstacles (legacy environments, dependency gaps) explicitly addressed.

*2.1.3 Labeling Dataset.* To build a reliable ground-truth dataset, we adopt a conservative, multi-stage pipeline that favors precision over recall. Every label is supported by verifiable evidence, yielding 111,025 high-confidence triples ⟨`binary`, `function_name`, `CVE-ID`⟩. The pipeline comprises three filters:

(1) **Symbol-based function identification.** We use `readelf` to parse each deduplicated binary's symbol table and extract function names. Although nearly 97% of binaries are stripped, many still retain some function names. These surviving symbols provide an unambiguous, high-fidelity signal for function name identity.

(2) **TPL version triangulation.** We apply curated regexes to string literals to detect and parse embedded version identifiers (e.g., "OpenSSL 1.1.1l"). To maintain high confidence, binaries without a parsable version string are excluded, since version-specific vulnerability assignment would otherwise be unreliable.

(3) **Vulnerability label generation.** We emit a label only when two conditions hold: (i) the symbol-derived function name matches a CVE's known vulnerable function, and (ii) the binary's detected version falls within that CVE's affected range.

Applying this pipeline to our firmware corpus yields a ground-truth set of substantial scale and diversity: 111,025 vulnerable binary functions across nine TPLs. This extensive, heterogeneous labeling ground truth provides a realistic basis for evaluating BCSD approaches in the real-world.

## 2.2 Vulnerability Detection and Evaluation

To rigorously assess the real-world efficacy of state-of-the-art BCSD approaches for large-scale vulnerability detection, we design a comprehensive experimental framework.

### *2.2.1 Experimental Setup.*

*Baseline approaches.* We evaluate four representative cross-architecture BCSD approaches spanning graph-based, execution-path, tree-structured, and IR-level semantic methods:

**GMN [27]** Models structural correspondences by aligning paired control-flow graphs (CFGs) with a graph neural network.

**Trex [37]** Captures compiler-invariant semantics by converting short symbolic-execution paths (micro-traces) into inputs for a hierarchical Transformer.

**Asteria [48]** Encodes syntactic structure using a Tree-LSTM over abstract syntax trees (ASTs) of disassembled functions.

**HermesSim [18]** Improves semantic robustness by lifting machine code to a normalized intermediate representation (IR) and comparing semantics-oriented graphs.

We use the official public implementations with default hyperparameters and no per-dataset tuning to promote fairness and reproducibility.

*Evaluation metrics.* We report two complementary retrieval metrics that together characterize ranking quality and practical search effectiveness:

**Mean Reciprocal Rank (MRR)** Quantifies how highly the first correct match is ranked; larger values indicate better overall precision.

**Recall@K** The fraction of queries whose correct match appears within the top-$K$ candidates, answering the practical question, "Will an analyst find the vulnerability among the first $K$ results?"

$$\text{MRR} = \frac{1}{|Q|}\sum_{i=1}^{|Q|}\frac{1}{\text{rank}_i}, \qquad \text{Recall@K} = \frac{1}{|Q|}\sum_{i=1}^{|Q|}\mathbb{I}\big(\text{rank}_i \le K\big) \tag{1}$$

where $|Q|$ is the number of queries, $\text{rank}_i$ is the rank of the true-positive match for the $i$-th query, and $\mathbb{I}(\cdot)$ is the indicator function.

*2.2.2 BCSD Approach Selection.* We train and evaluate all four baselines on the *stripped* BinKit dataset using BinKit's standard train/test split and three rigorous scenarios: **XC** (cross-compiler),

**XA** (cross-architecture), and **XM** (mixed). Unless otherwise stated, we use official public implementations with default hyperparameters and avoid per-dataset tuning to ensure fairness and reproducibility.

As shown in Figure 3, **HermesSim** consistently outperforms the other systems. Its MRR exceeds the next-best system (Asteria) by 8.7 percentage points in XA and 25.5 in XC, while also achieving the highest Recall@1. In the most challenging XM setting, HermesSim again leads with an MRR of 74.7%. Given its decisive advantage in both ranking quality and top-1 accuracy, we adopt **HermesSim** as the core engine for our large-scale firmware vulnerability analysis.

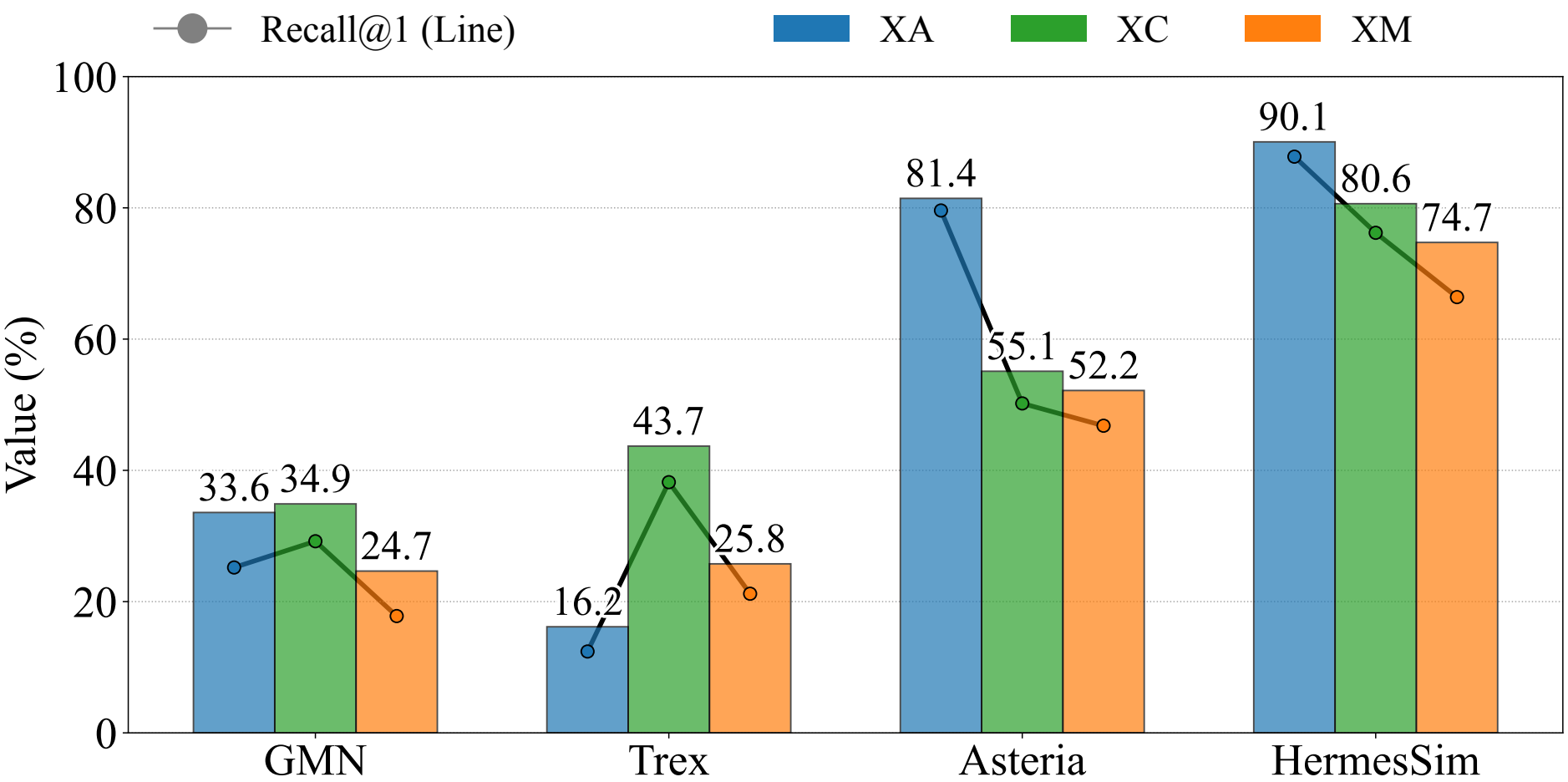

Fig. 3. Performance on the BinKit benchmark (trained and evaluated on BinKit using its standard split). Bars show MRR; points show Recall@1. HermesSim consistently leads across XA, XC, and XM.

*2.2.3 Detection and Evaluation Workflow.* Building on this selection, we run a large-scale retrieval, and assess performance against our ground truth and research questions.

(1) **Lifting and feature extraction.** We disassemble binaries with **IDA Pro** and **Ghidra** and extract *semantics-oriented graphs (SOGs)* as required by HermesSim. Each function is encoded by the HermesSim encoder into a fixed-length embedding, producing one vector per function for both the vulnerable function pool and the firmware function pool.
(2) **Score and ranking.** For each labeled vulnerable function, we compute the *cosine similarity* between its embedding and all firmware-function embeddings. Firmware candidates are then ranked in descending order of similarity to form a per-query result list.
(3) **Analysis.** Using the *labeling dataset* constructed earlier as ground truth, we compute *MRR* and *Recall@K* as in Eq. (1). These metrics drive our study of four research questions.

*2.2.4 Research Questions.* This study aims to answer the following research questions.

**RQ1: How do different versions of vulnerable functions impact the BCSD performance?** A single vulnerability often spans multiple TPL versions (e.g., CVE-2014-0160 [34] impacts OpenSSL from 1.0.1 through 1.0.1f). Since firmware may embed TPLs in arbitrary versions, these may not correspond precisely to the versions used to construct the vulnerable functions for queries. Such version discrepancies can cause mismatches in vulnerable functions due to code modifications across versions. This RQ aims to evaluate the impact of different versions of vulnerable functions on vulnerability detection.

**RQ2: How do different search spaces impact the BCSD performance?** The selection of search space exerts a direct influence on both the recall and precision of vulnerability detection. A broader search space enhances the likelihood of capturing vulnerable functions but simultaneously introduces substantial noise and computational overhead, whereas a narrower search space improves efficiency yet risks overlooking real vulnerabilities. Consequently, understanding the impact of different search space configurations on detection performance is crucial for striking an effective balance among recall, accuracy, and computational cost in practical scenarios. This RQ aims to evaluate the impact of search spaces on vulnerability detection.

**RQ3: How does different function size impact the BCSD performance?** Function size can significantly influence the effectiveness of vulnerability detection based on BCSD. Smaller functions may lack sufficient semantic context, leading to false negatives, whereas larger functions may introduce redundant or noisy code that hampers accurate matching. This RQ aims to evaluate the impact of function sizes on vulnerability detection.

**RQ4: How do variations in compilation toolchains impact the BCSD performance?** Different compilation toolchains can substantially reshape the structure and representation of binary code, even when generated from identical source code. Such variations may alter control flow, instruction patterns, and function boundaries, thereby influencing the effectiveness of vulnerability detection methods in accurately identifying vulnerable functions. Prior studies [18, 44, 48] have typically focused on compilers, optimization levels, and architectures, compiling binaries under idealized conditions. In reality, however, compilation toolchains are far more complex. This RQ aims to assess the impact of diverse compilation toolchains on vulnerability detection and to identify overlooked compilation factors that may significantly affect BCSD performance.

## 3 Evaluation Results and Findings

We now present the results of our large-scale experiment, organized by the research questions defined in Section 2.2.4.

### 3.1 RQ1: How do different versions of vulnerable functions impact the BCSD performance?

*Experimental Setup.* To evaluate BCSD's sensitivity to vulnerable function versions, we prepare three distinct versions of vulnerable functions as queries for vulnerability detection, detailed as follows.

- $V_{same}$: Versions of vulnerable functions that exactly match the TPL versions used in firmware, establishing a baseline with no version discrepancies between the query and target vulnerable functions.
- $V_{oldest}$: The oldest versions of vulnerable functions within the specified affected range of a vulnerability (e.g., 1.0.1 from [1.0.1, 1.0.1f]).
- $V_{newest}$: The newest versions of vulnerable functions within the specified affected range of a vulnerability (e.g., 1.0.1f from [1.0.1, 1.0.1f]).

*Experimental Results.* Figure 4 presents the vulnerability detection results across various TPLs, along with the aggregated overall performance. The x-axis represents the different TPLs, while the y-axis indicates the MRR scores.

The results clearly demonstrate that the versions of vulnerable functions have a substantial impact on BCSD performance. Using $V_{same}$ as queries consistently delivers the best outcomes across all TPLs. On average, $V_{same}$ achieves an MRR of 0.961, surpassing $V_{oldest}$ and $V_{newest}$ by 11.1% and 4.6%, respectively. $V_{oldest}$ performs the worst, with sharp declines observed for TPLs such as FreeType, libpng, and OpenSSL. $V_{newest}$ attains moderate performance, occasionally outperforming the $V_{oldest}$ baseline but still falling short of $V_{same}$. Specifically, Binutils and BusyBox consistently demonstrate identical and high performance across all three version types, with MRR values approaching 1.0. Manual analysis of these two TPLs in binaries reveals that the vulnerable functions undergo relatively minor changes across different releases, thereby preserving stable function semantics. This observation aligns with practical reality, as both TPLs primarily provide basic functionalities and are seldom updated by developers.

Overall, the results indicate that mismatches between the query versions of vulnerable functions and the target versions in binaries will impair detection effectiveness.

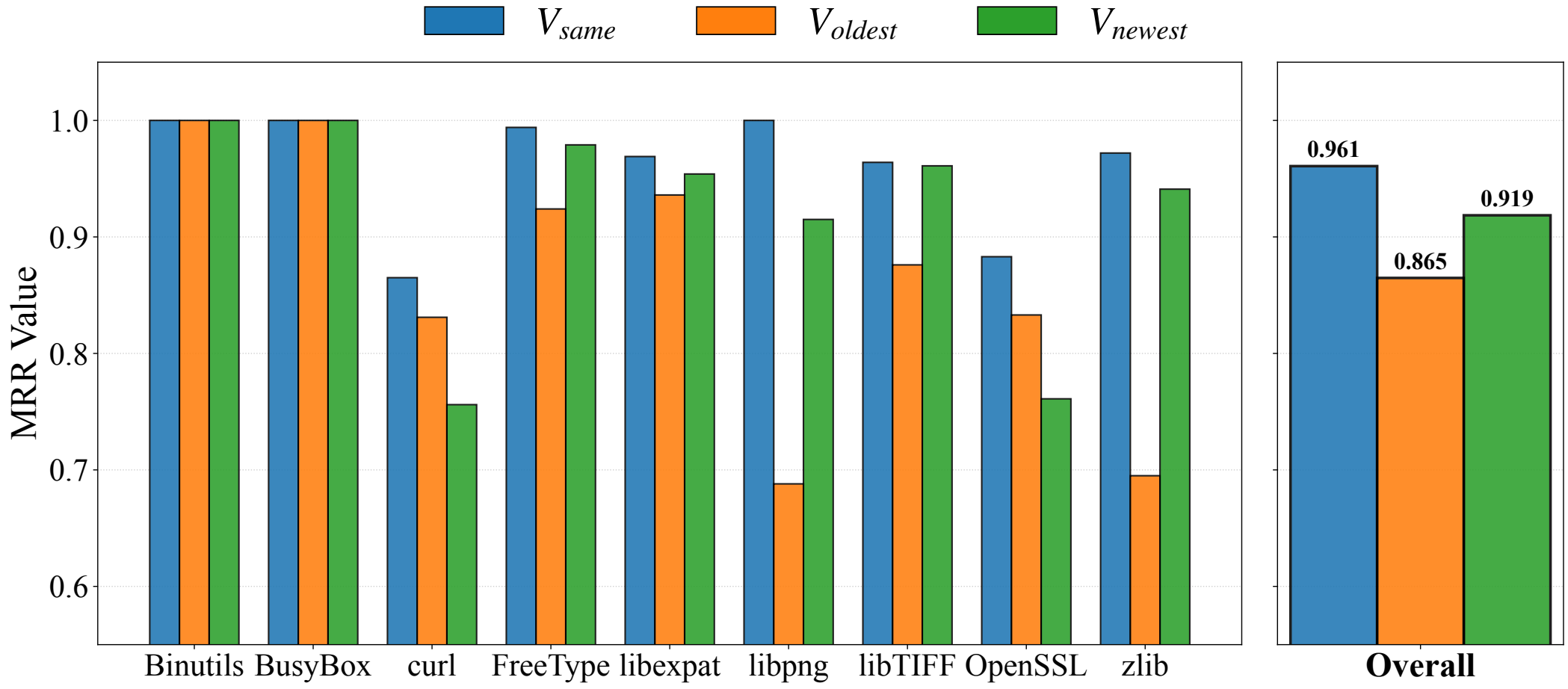

Fig. 4. Impact of different versions of vulnerable functions on BCSD performance.

**Answer to RQ1.**

► The versions of vulnerable functions have a significant impact on BCSD performance. Matching queries with the same versions as the targets achieves a high MRR of 0.961, while using mismatched versions reduces performance by 4.6% to 11.1%.

## 3.2 RQ2: How do different search spaces impact the BCSD performance?

*Experimental Setup.* To investigate how different search spaces impact BCSD performance, we designed two comparative settings.

- **Agnostic.** In this setting, the entire firmware is treated as the search space, with vulnerable functions retrieved on a global scale.
- **Aware.** In this setting, the search space is restricted to the corresponding TPL, with detection conducted solely within the targeted TPL, a subspace of the broader agnostic search.

Table 3. Impact of search spaces on libpng and OpenSSL.

| TPL | Search | Recall@1 | MRR |
|---|---|---|---|
| libpng | Agnostic | 0.53 | 0.751 |
| | **Aware** | **1.000** | **1.000** |
| OpenSSL | Agnostic | 0.794 | 0.838 |
| | **Aware** | **0.870** | **0.883** |

*Experimental Results.* Table 3 presents the impact of search spaces across different TPLs. For libpng, aware search achieved perfect performance with Recall@1 and MRR both reaching 1.0, compared to 0.53 and 0.751 in the agnostic case. For OpenSSL, aware search also produced consistent gains, improving Recall@1 from 0.794 to 0.870 and MRR from 0.838 to 0.883. These results clearly demonstrate that restricting the search space to the correct TPL substantially reduces noise from irrelevant functions, thereby improving both retrieval accuracy and ranking quality.

Overall, the experiments confirm that search space is a critical factor in BCSD performance. While agnostic search offers broader coverage, it suffers from reduced precision due to the larger candidate pool. Incorporating TPL-awareness consistently enhances detection performance, underscoring its practical value in real-world vulnerability detection scenarios.

**Answer to RQ2**

► Restricting the search space to the correct TPL markedly enhances BCSD performance compared to agnostic search across the entire firmware. On average, the MRR score rises by 18.5%, while Recall1 improves by 41.2%.

### 3.3 RQ3: How does different function size impact the BCSD performance?

*Experimental Setup.* To evaluate the impact of function size on BCSD performance, we first categorize functions according to their sizes, calculated by subtracting the start address from the end address of each function. We first record the number of functions across different size ranges in our dataset, and then compute the corresponding MRR for each group to illustrate performance variations across function sizes.

*Experimental Results.* Figure 5a illustrates the distribution of function sizes in our dataset, where functions under 1KB account for the majority, while those exceeding 4KB are comparatively uncommon. Figure 5b presents the vulnerability detection results across different function size ranges. For clearer illustration, we categorize function sizes into three groups: small functions (< 2KB), medium functions (2–4KB), and large functions (> 4KB).

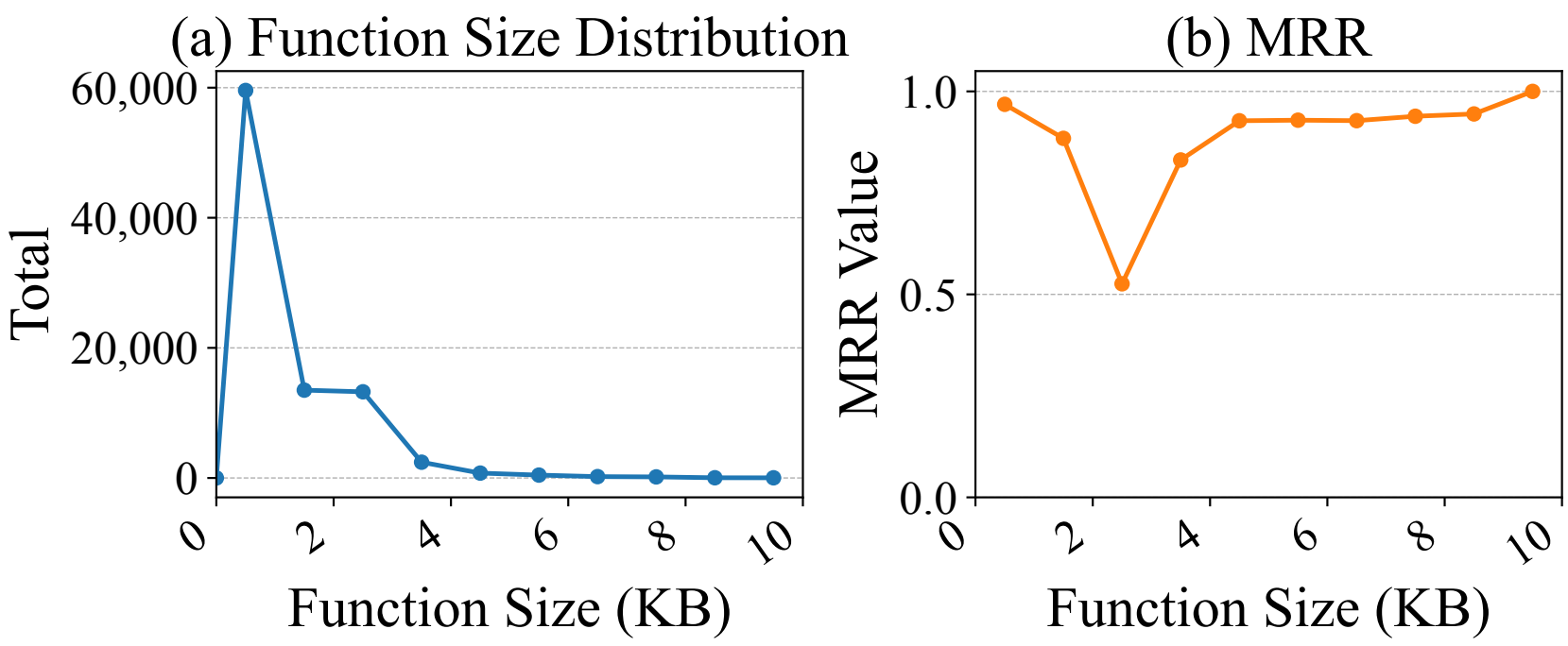

Fig. 5. Non-linear impact of function size on BCSD performance. (a) Long-tailed size distribution in our dataset. (b) U-shaped relationship between size and MRR, with a pronounced performance trough at mid sizes.

- **Small functions (< 2 KB).** Performance is very strong, forming the left arm of the U. Sub-1 KB functions reach an MRR of 0.98; the 1–2 KB bracket remains high at 0.88. These functions tend to have simple, canonical structures with limited room for disruptive optimizations, yielding stable, easily matched features.
- **Mid-sized functions (2–4 KB).** Performance dips into a *complexity trough.* MRR drops from 0.88 to a nadir of 0.53 in the 2–3 KB range (a 35-point decline), then begins to recover to 0.84 for 3–4 KB. These functions are large enough for compilation to meaningfully restructure control and data flow, yet often lack uniquely identifying cues, making similarity less robust.

- **Large functions (> 4 KB).** Performance rebounds to form the right arm of the U: MRR exceeds 0.93 and approaches unity near 10 KB. Here, rich idiosyncratic signals—e.g., intricate CFGs, distinctive string literals, or sizable constant arrays—act as resilient fingerprints across builds.

**Answer to RQ3**

▶ Function size also plays a crucial role in BCSD performance. Detection is consistently strong for small and large functions, but medium-sized functions introduce complexity that leads to reduced accuracy.

## 3.4 RQ4: How do variations in compilation toolchains impact the BCSD performance?

A common assumption in many BCSD studies is that a single, cleanly compiled binary can serve as a universal query. We revisit this assumption by examining heterogeneous, real-world build environments typical of firmware.

*3.4.1 The Problem: Pervasive Performance Variance.* To isolate compilation effects, we ran a controlled study: OpenSSL binaries from 42 device vendors (same architecture/bitness). For each target, we issued an *oracle* query whose function matched version, architecture, and optimization level exactly. Thus, any residual variance reflects vendor toolchains and build configues.

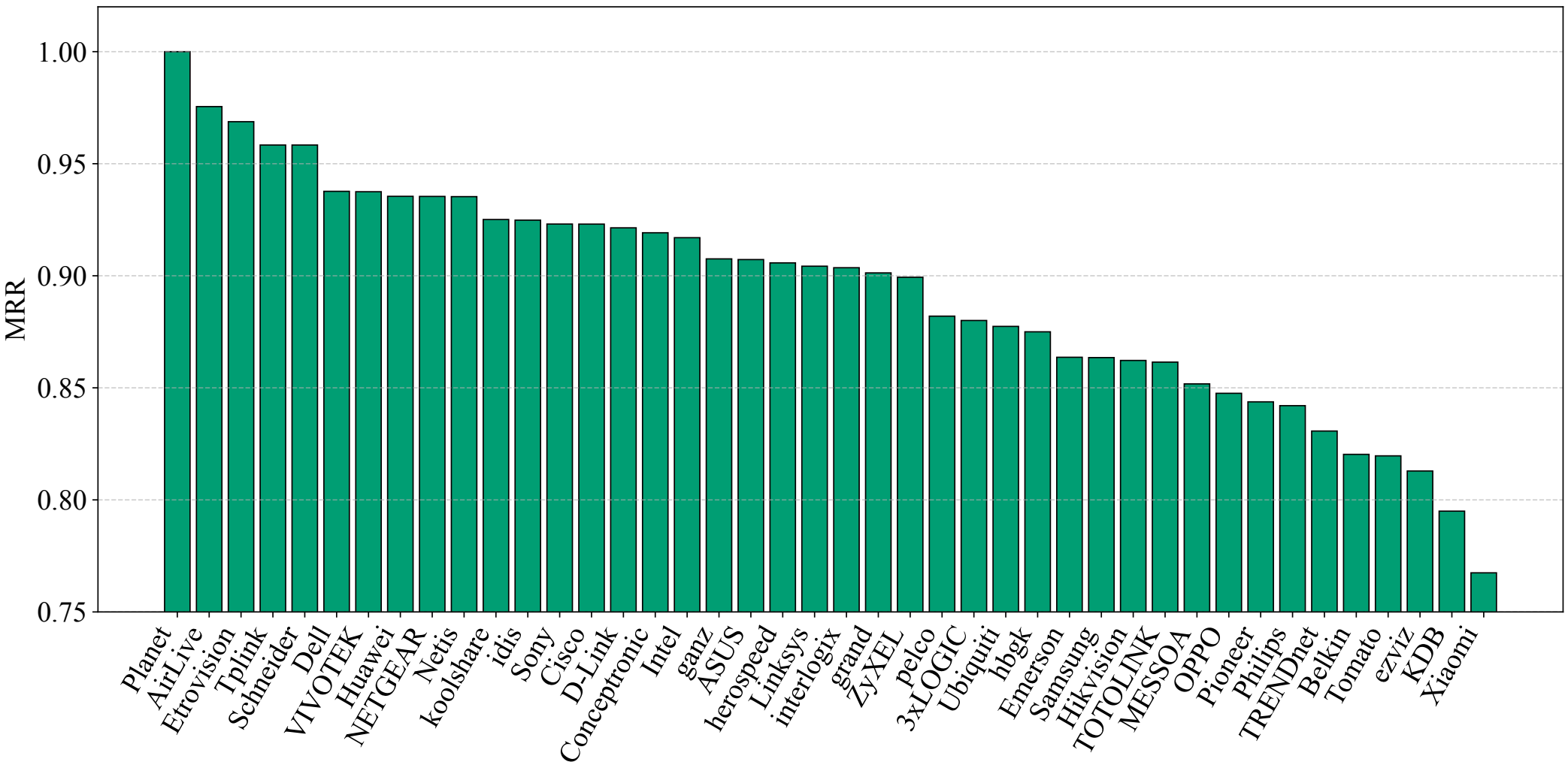

Fig. 6. BCSD performance across vendor-specific OpenSSL builds. Even with version/architecture/optimization controlled, MRR varies widely, indicating substantial sensitivity to vendor toolchains and configurations.

The results, shown in Figure 6, reveal a startling reality: even under these ideal conditions, detection performance is highly inconsistent. The Mean Reciprocal Rank (MRR) spans a wide spectrum, from a perfect 1.0 for vendors like Planet and AirLive down to a fragile 0.768 for Xiaomi. The aggregate MRR of 0.896 masks this dangerous variance; performance for nearly half the vendors, including prominent names like Belkin and Tomato, falls well below this average. This demonstrates that a "semantic gap" exists between binaries compiled from identical source, a gap wide enough to render even oracle-based queries unreliable.

*3.4.2 The Cause: A Diverse Ecosystem of Build Configurations.* This performance variance is a direct consequence of the diverse compilation ecosystem in firmware. Our analysis identified 168 distinct compiler toolchains across our dataset, often dictated by SDKs from chipset manufacturers (e.g., Broadcom, HiSilicon) or community projects like OpenWrt.

At a granular level, this heterogeneity is driven by the differential application of compilation macros, which can fundamentally alter binary structure, as categorized in **Table 4**. These macros, ranging from enabling assembly-level implementations (`AES_ASM`) to stripping entire features (`OPENSSL_NO_ERR`), are the root cause of the semantic gap, posing a critical challenge for any BCSD technique that relies on a single, "vanilla" build for its queries.

Table 4. Taxonomy of Common Compilation Macros and Their Impact on Binary Structure.

| Impact Level | Description of Change | Example Macros |
|---|---|---|
| **High** | **Implementation/ISA:** Replaces high-level language implementation with low-level assembly for core routines. | `AES_ASM`, `SHA1_ASM`, `OPENSSL_BN_ASM_*` |
| **High** | **Functional:** Adds or removes entire features, modules, or protocols, fundamentally altering the program's capabilities and CFG. | `OPENSSL_NO_ERR`, `PROXYARP`, `BCMWPA2` |
| **Medium** | **Structural:** Alters the binary's memory organization, addressing model, or concurrency model; Modifies data representation. | `OPENSSL_THREADS/_REENTRANT;`, `L_ENDIAN` |
| **Low** | **Configurational:** Toggles lightweight, non-structural code elements like diagnostics or assertions, without changing core algorithms. | `DEBUG_*`, `NDEBUG`, `ENABLE_LOGGING,OPENSSL_SMALL_FOOTPRINT` |

*3.4.3 The Solution: Build-Aware Query Generation.* To address this challenge, we propose a "build-aware" query strategy that sources queries from binaries representative of real-world build patterns rather than from a default compilation. To implement this, we developed a methodology to systematically identify these build archetypes. We first extracted compilation fingerprints (toolchain strings and macro settings) from our firmware binary dataset. Focusing on the prevalent Broadcom ARM toolchain family, we then clustered binaries based on their macro configurations. This revealed several dominant clusters, each representing a common, real-world build archetype. Binaries from these prominent clusters were subsequently used to generate our "In-the-Wild" queries.

**Default-Compile Query** The conventional approach, where a vulnerable function is compiled from source using default settings (e.g., `gcc -O3`).

**In-the-Wild Query** Our proposed approach, which sources queries from representative build archetypes identified in the firmware ecosystem (i.e., the dominant clusters of the Broadcom ARM toolchain family).

Table 5. Performance of Default vs. In-the-Wild Queries on a Hold-Out (Non-Broadcom) Dataset.

| Impact Level | Sub-category | Recall@1 | | | MRR | | | Proportion (%) |
|---|---|---|---|---|---|---|---|---|
| | | W | D | Δ(W-D) | W | D | Δ(W-D) | |
| **High** | Functional | **0.966** | 0.789 | 0.177 | **0.974** | 0.811 | 0.163 | 39.71 |
| | Implementation/ISA | **0.974** | 0.808 | 0.166 | **0.986** | 0.818 | 0.168 | 56.41 |
| **Medium** | Structural/Model | **0.959** | 0.741 | 0.218 | **0.970** | 0.785 | 0.185 | 1.14 |
| **Low** | Configurational | **1.000** | 0.870 | 0.130 | **1.000** | 0.897 | 0.103 | 0.11 |
| **Unknown** | | **0.969** | 0.753 | 0.216 | **0.976** | 0.773 | 0.203 | 2.62 |

**Note: W** = In-the-Wild; **D** = Default. Proportions are based on a total of 62,128 items.

We evaluated both query strategies against a hold-out set of 58,511 vulnerable functions from all non-Broadcom toolchains. The results, summarized in **Table 5**, are decisive. The In-the-Wild

Query demonstrates markedly superior generalization. For instance, across high-impact categories, it improves the aggregate MRR from 0.818 to 0.986—a substantial increase of 16.8 percentage points. This performance leap confirms that queries sourced from real-world builds are far more representative and robust, effectively bridging the semantic gap.

*3.4.4 Case Study: Visualizing the Semantic Gap.* To illustrate *why* the default-compile query fails, **Figure 7** compares the Control-Flow Graph (CFG) of OpenSSL's `ASN1_verify` function under two build configurations. The default compilation (a) features a distributed error-handling architecture. In stark contrast, the in-the-wild version (b) is transformed by "High Impact" macros from Table 4, like `OPENSSL_NO_ERR` and `OPENSSL_SMALL_FOOTPRINT`. These macros prune detailed error reporting and consolidate error handlers into a single block. The resulting binaries are so topologically dissimilar that a query from the default build cannot reliably match its hardened, real-world counterpart, providing a clear visual explanation for the performance gap.

Fig. 7. CFG Divergence in OpenSSL's `ASN1_verify` due to Compilation Macros. (a) The default compilation shows complex, distributed error handling. (b) An in-the-wild build, hardened with macros like `OPENSSL_NO_ERR`, shows a radically simplified and consolidated structure, creating the semantic gap.

**Answer to RQ4**

► Even with version/arch/opt fixed, vendor macro config reshape functions, with an MRR distribution spanning 1.000–0.768. Switching to build-aware, in-the-wild queries lifts overall MRR 0.814 ⟶ 0.981 (+16.7) and Recall@1 0.798 ⟶ 0.971 (+17.2), effectively closing this gap.

## 4 Discussion

We revisit a central question: *what is required for BCSD to remain reliable at firmware scale?* Across RQ1–RQ4, our results indicate that reliability is not determined by a single model choice, but by aligning three levers: (i) representative query construction, (ii) early and accurate search-space scoping, and (iii) robustness to size- and build-induced variance.We also highlight that each lever introduces non-trivial manual or engineering effort whose costs should be weighed against the accuracy gains it provides.

### 4.1 Key Insights Across RQs

**Representative queries over canonical builds.** RQ4 shows that a clean, default-compiled query is brittle in heterogeneous vendor environments. Even when version, architecture, and optimization level match, vendor toolchains and macro toggles create a *semantic gap* that degrades retrieval. Selecting queries from *in-the-wild* build archetypes improves generalization and lifts MRR substantially (Finding 4). In practice, a small portfolio of archetype-aligned queries per vulnerable function is preferable to a single canonical build,although curating and validating such a portfolio requires meaningful manual effort.

**Scope first, then rank.** RQ2 demonstrates that *TPL-aware* scoping yields consistent gains over brute force by cutting noise and reducing compute (Finding 2).Identifying the correct TPL before matching improves recall@1 and MRR,and also stabilizes downstream triage since fewer non-target candidates enter the ranking stage.We note that accurate scoping itself depends on the quality of metadata or the performance of SCA tools, neither of which is always guaranteed in practice.

**Function size is non-monotonic risk.** RQ3 reveals a U-shaped size–performance curve with a pronounced trough at 2–3 KB (Finding 3). Small functions are stable and distinctive enough for high accuracy; very large functions carry resilient fingerprints. Mid-sized functions sit in an instability region where compiler rewrites are common and unique anchors are scarce. These cases benefit the most from contextual features and multi-query fallback.

### 4.2 Implications for Deployment

- **Build-aware queries.** Replace a single default-compiled query with a compact portfolio drawn from representative in-the-wild build archetypes.Pick queries that cover common differences such as toolchain, libc, optimization, and macro.Start with a small set that gives the best coverage, drop near-duplicates,and adjust the portfolio size based on observed retrieval performance.Refresh the set if you notice more misses for a specific vendor or family.
- **Scope first, rank late.** Use TPL/SBOM-driven scoping to shrink candidates before matching; then fuse ranks across the small archetype-aligned query set to stabilize precision and cost. First, use whatever hints you have (SBOM, version strings, symbols, file paths) to narrow the library list. Spend more compute on the most likely libraries and stop early when one result clearly stands out. Combine the few query results with a simple average or vote.
- **Focus on mid-sized hard cases.** Treat 2–3 KB functions as high-Challenge: raise decision thresholds, add lightweight caller/callee context, and enable targeted multi-query fallback on uncertainty. For these functions, ask for stronger evidence before accepting a match. Add a bit of context (who calls it, notable constants/strings) to steady the result. If still unsure, try one or two extra query variants and cap the extra work.

### 4.3 Limitations

**Library and domain coverage.** While we study widely deployed libraries (e.g., OpenSSL, zlib), broader coverage is needed to confirm external validity. Our dataset does not fully cover proprietary vendor stacks, or RTOS ecosystems, so results may not transfer perfectly. We plan to add more families and architectures to reduce this gap.

**Macro inference noise.** Macro settings are inferred from static artefacts (symbols, strings); stripped binaries and aggressive LTO can obscure these signals.In practice, the extraction relies on regular expressions, which are inherently fragile—unusual formatting or vendor-specific naming can cause silent mis-extractions, and maintaining these patterns incurs ongoing manual effort as target firmware diversifies.When signals are missing or unreliable, our guesses can be wrong and some matches may be missed.

**Function boundary fidelity.** Inlining and compiler-specific prologues may introduce boundary errors that affect both size binning (RQ3) and per-function matching. Such errors can push functions into the wrong size bucket or distort features. We use several boundary detectors and simple smoothing, but some residual error remains.

### 4.4 Future Work

**Context-augmented representations.** Add stable extrinsic signals (who calls the function, what it calls, notable constants/strings) to help especially for mid-sized cases. Keep the context small so lookup stays fast. We will measure the accuracy–latency trade-off directly.

**Learned query augmentation.** Train transformations that mimic common compiler and macro changes, so the system can automatically create a few extra query variants. The goal is to boost recall with only 2–4 added variants per function. We will select them by simple held-out gains to control cost.

**Uncertainty- and cost-aware search.** Calibrate scores so the system can say "unsure" explicitly. Only then expand to extra queries or add a bit of context, and stop early once a clear winner appears. Cap the extra work and route low-confidence cases to a quick human check.

## 5 Related Work

Research in firmware vulnerability discovery primarily follows two paradigms: Binary Code Similarity Detection (BCSD) for identifying known N-day vulnerabilities, and non-BCSD methods such as static or dynamic analysis for uncovering novel 0-day bugs. While our work aligns with the former, we review both to contextualize our contributions.

### 5.1 BCSD-Based Vulnerability Detection

Operating on a "search-and-match" principle, this paradigm leverages a known vulnerable function to find similar instances across a large binary corpus, making it highly effective for scalable N-day detection in the fragmented firmware ecosystem. The field has evolved from graph-based matching to more robust deep learning-based embeddings.

**Graph-based Matching.** Early approaches represented functions as control-flow graphs (CFGs), measuring similarity through graph isomorphism. Seminal works like Genius [11] clustered functions by CFG features, while BinDiff [16] remains a classic tool for differential analysis. To handle architectural diversity, discovRE [9] introduced value-driven matching. However, these methods are computationally intensive and brittle against compiler variations that alter graph structures.

**Embedding-based Similarity.** To overcome the rigidity of graph matching, subsequent research embraced high-dimensional vector representations (embeddings), enabling more nuanced and scalable similarity comparisons. Gemini [47], a landmark work, applied graph embedding networks to Attributed CFGs (ACFGs), achieving remarkable cross-architecture performance. This inspired a new wave of deep learning models. SAFE [32] employed a self-attentive network, while NLP-inspired techniques like Asm2Vec [8] and PalmTree [26] learned semantics from instruction sequences or pseudocode.*Recent technical trends (roughly 2023–2025) [18, 23, 39, 42] emphasize*: (i) semantics-oriented encodings that normalize away syntax and highlight data/control dependencies, often via lightweight IRs; (ii) resilience to compiler effects by explicitly handling inlining and fuzzy function boundaries with multi-granularity matching or code extraction; and (iii) system-level scaling, using compact embeddings, approximate nearest-neighbor indexes, and simple rank fusion with calibrated scores for high-precision triage.

In summary, the evolution of BCSD towards capturing deeper code semantics has established it as the backbone for scalable N-day vulnerability detection, offering greater accuracy and resilience to compiler variations.

### 5.2 Non-BCSD Approaches for Firmware Analysis

In contrast to similarity-based methods, this paradigm seeks to uncover vulnerabilities *de novo* by analyzing firmware for specific bug patterns or anomalous behaviors. Static analysis inspects binaries without execution. For example, Firmalice [3] used symbolic execution to find authentication bypasses. Dynamic analysis and fuzzing execute firmware in an emulator to trigger bugs. Firmadyne [2] pioneered whole-system emulation, and Avatar [41] introduced a hybrid approach to bridge emulation and physical hardware. Beyond these early systems, later work operationalizes taint tracking and IR lifting to trace data flows from network/peripheral inputs to security-sensitive sinks, e.g., inter-binary taint propagation and configuration extraction in bare-metal images [38, 45]. Practical pipelines couple function identification with rule-based detectors to surface common issues in embedded web stacks (e.g., command injection in CGI handlers, insecure update paths, weak authentication flows) [5, 12, 46]. On the dynamic side, coverage-guided fuzzers integrated with rehosting and snapshotting skip lengthy boots and stress high-value services [10, 20, 40, 52]. Toolchains increasingly support differential analysis across firmware versions/models to catch regressions and removed checks during maintenance [12]. Hardware-in-the-loop variants drive real devices via UART/JTAG/SWD alongside emulator traces to improve oracle quality and crash triage [25, 33, 50].

**Inherent Limitations.** However, these methods are hampered by significant challenges in the firmware context. Static analysis suffers from high false positives due to path explosion and a lack of hardware context. Dynamic analysis is constrained by the "emulation gap"—the difficulty of modeling hardware-software interactions (e.g., MMIO), which leaves large code sections untested. Crucially, both approaches face scalability bottlenecks, rendering them ill-suited for analyzing the vast number of firmware images in the wild.

While indispensable for 0-day discovery, these scalability and fidelity challenges render non-BCSD approaches less practical for large-scale analysis. This positions BCSD as the more effective paradigm for systematic N-day vulnerability campaigns, motivating the focus of our work.

## 6 Conclusion

In this paper, we conduct the largest-scale empirical study on the methodology of BCSD for firmware vulnerability detection. Our work demonstrates that conventional query strategies are fundamentally flawed and establishes that a build-aware methodology is the key to making BCSD effective in the wild. Using this approach on a dataset of nearly 60,000 firmware images, we boost detection MRR from 0.818 to a near-perfect 0.981. Our analysis also quantifies the necessity of a TPL-aware search pipeline and uncovers a "complexity trough" where mid-sized functions are hardest to detect. We believe this work provides a validated, data-driven roadmap for operationalizing BCSD at scale and challenges the research community to adopt more realistic evaluation benchmarks.

## 7 Data Availability

Research artifacts and experimental data are available at https://github.com/Guo-jingdong/Firmware-and-Vulnerability-dataset.

## References

[1] Ons Aouedi, Thai-Hoc Vu, Alessio Sacco, Dinh C. Nguyen, Kandaraj Piamrat, Guido Marchetto, and Quoc-Viet Pham. 2024. A Survey on Intelligent Internet of Things: Applications, Security, Privacy, and Future Directions. *IEEE Communications Surveys & Tutorials* (2024). doi:10.1109/COMST.2024.3430368

[2] BusyBox. 2025. BusyBox: The Swiss Army Knife of Embedded Linux. https://www.busybox.net/.

[3] Daming D. Chen, Manuel Egele, Maverick Woo, and David Brumley. 2016. Towards Automated Dynamic Analysis for Linux-based Embedded Firmware. In *Proceedings of the 23rd Network and Distributed System Security Symposium*

*(NDSS)*. Internet Society, San Diego, CA, USA. doi:10.14722/ndss.2016.23415
[4] Andrei Costin, Jonas Zaddach, Aurélien Francillon, and Davide Balzarotti. 2014. A large-scale analysis of the security of embedded firmwares. In *Proceedings of the 23rd USENIX Conference on Security Symposium* (San Diego, CA) *(SEC'14)*. USENIX Association, USA, 95–110.
[5] Andrei Costin, Apostolis Zarras, and Aurélien Francillon. 2016. Automated Dynamic Firmware Analysis at Scale: A Case Study on Embedded Web Interfaces. In *Proceedings of the 11th ACM on Asia Conference on Computer and Communications Security* (Xi'an, China) *(ASIA CCS '16)*. Association for Computing Machinery, New York, NY, USA, 437–448. doi:10.1145/2897845.2897900
[6] curl. 2025. curl: Command line tool and library for transferring data with URLs. https://curl.se/.
[7] Yaniv David and Eran Yahav. 2014. Tracelet-based code search in executables. In *Proceedings of the 35th ACM SIGPLAN Conference on Programming Language Design and Implementation* (Edinburgh, United Kingdom) *(PLDI '14)*. Association for Computing Machinery, New York, NY, USA, 349–360. doi:10.1145/2594291.2594343
[8] Steven H. H. Ding, Benjamin C. M. Fung, and Philippe Charland. 2019. Asm2Vec: Boosting Static Representation Robustness for Binary Clone Search against Code Obfuscation and Compiler Optimization. In *2019 IEEE Symposium on Security and Privacy (SP)*. 472–489. doi:10.1109/SP.2019.00003
[9] Sebastian Eschweiler, Khaled Yakdan, and Elmar Gerhards-Padilla. 2016. discovRE: Efficient Cross-Architecture Identification of Bugs in Binary Code. In *Network and Distributed System Security Symposium*. doi:10.14722/ndss.2016.23185
[10] Bo Feng, Alejandro Mera, and Long Lu. 2020. $P^2$IM: scalable and hardware-independent firmware testing via automatic peripheral interface modeling. In *Proceedings of the 29th USENIX Conference on Security Symposium (SEC'20)*. USENIX Association, USA, Article 70, 18 pages. https://www.usenix.org/conference/usenixsecurity20/presentation/feng
[11] Qian Feng, Rundong Zhou, Chengcheng Xu, Yao Cheng, Brian Testa, and Heng Yin. 2016. Scalable Graph-based Bug Search for Firmware Images. In *Proceedings of the 2016 ACM SIGSAC Conference on Computer and Communications Security* (Vienna, Austria) *(CCS '16)*. Association for Computing Machinery, New York, NY, USA, 480–491. doi:10.1145/2976749.2978370
[12] Fraunhofer SIT. 2019. FACT – Firmware Analysis and Comparison Tool: Documentation and Comparison Capabilities. https://fact-firmware-analysis.readthedocs.io/. Accessed 2025-09-12.
[13] FreeType. 2025. FreeType: A Free, High-Quality and Portable Font Engine. https://freetype.org/.
[14] Jian Gao, Xin Yang, Ying Fu, Yu Jiang, and Jiaguang Sun. 2018. VulSeeker: a semantic learning based vulnerability seeker for cross-platform binary. In *Proceedings of the 33rd ACM/IEEE International Conference on Automated Software Engineering* (Montpellier, France) *(ASE '18)*. Association for Computing Machinery, New York, NY, USA, 896–899. doi:10.1145/3238147.3240480
[15] GNU Project. 2025. GNU Binutils. https://www.gnu.org/software/binutils/.
[16] Google. 2011. BinDiff. https://www.zynamics.com/bindiff.html.
[17] Irfan Ul Haq and Juan Caballero. 2021. A Survey of Binary Code Similarity. *ACM Comput. Surv.* 54, 3, Article 51 (April 2021), 38 pages. doi:10.1145/3446371
[18] Haojie He, Xingwei Lin, Ziang Weng, Ruijie Zhao, Shuitao Gan, Libo Chen, Yuede Ji, Jiashui Wang, and Zhi Xue. 2024. Code is not Natural Language: Unlock the Power of Semantics-Oriented Graph Representation for Binary Code Similarity Detection. In *Proceedings of the 33rd USENIX Security Symposium (USENIX Security 24)*. USENIX Association, Philadelphia, PA, 1759–1776. https://www.usenix.org/conference/usenixsecurity24/presentation/he-haojie
[19] Craig Heffner and ReFirm Labs. 2025. *Binwalk: Firmware Analysis Tool*. https://github.com/ReFirmLabs/binwalk
[20] Grant Hernandez, Dave Jing Tian, Tuba Yavuz, Caroline Trippel, Kevin Butler, et al. 2022. FIRMWIRE: Transparent Dynamic Analysis for Cellular Baseband Firmware. In *Network and Distributed System Security Symposium (NDSS)*. https://www.ndss-symposium.org/wp-content/uploads/2022-136-paper.pdf
[21] IBM. 2020. A new botnet attack just mozied into town. https://www.ibm.com/think/x-force/botnet-attack-mozi-mozied-into-town.
[22] IBM. 2024. Firmware vs. software: What's the difference and why it matters. https://www.ibm.com/think/insights/firmware-vs-software.
[23] Lichen Jia, Chenggang Wu, Peihua Zhang, and Zhe Wang. 2024. CodeExtract: Enhancing Binary Code Similarity Detection with Code Extraction Techniques. In *Proceedings of the 25th ACM SIGPLAN/SIGBED International Conference on Languages, Compilers, and Tools for Embedded Systems* (Copenhagen, Denmark) *(LCTES 2024)*. Association for Computing Machinery, New York, NY, USA, 143–154. doi:10.1145/3652032.3657572
[24] Dongkwan Kim, Eunsoo Kim, Sang Kil Cha, Sooel Son, and Yongdae Kim. 2023. Revisiting Binary Code Similarity Analysis Using Interpretable Feature Engineering and Lessons Learned. *IEEE Transactions on Software Engineering* 49, 4 (2023), 1661–1682. doi:10.1109/TSE.2022.3187689
[25] Wenqiang Li, Jiameng Shi, Fengjun Li, Jingqiang Lin, Wei Wang, and Le Guan. 2022. $\mu AFL$: Non-intrusive Feedback-driven Fuzzing for Microcontroller Firmware. In *2022 IEEE/ACM 44th International Conference on Software Engineering*

*(ICSE)*. 1–12. doi:10.1145/3510003.3510208

[26] Xuezixiang Li, Yu Qu, and Heng Yin. 2021. PalmTree: Learning an Assembly Language Model for Instruction Embedding. In *Proceedings of the 2021 ACM SIGSAC Conference on Computer and Communications Security (CCS '21)*. Association for Computing Machinery, New York, NY, USA, 3236–3251. doi:10.1145/3460120.3484587

[27] Yujia Li, Chenjie Gu, Thomas Dullien, Oriol Vinyals, and Pushmeet Kohli. 2019. Graph Matching Networks for Learning the Similarity of Graph Structured Objects. In *Proceedings of the 36th International Conference on Machine Learning (Proceedings of Machine Learning Research, Vol. 97)*, Kamalika Chaudhuri and Ruslan Salakhutdinov (Eds.). PMLR, 3835–3845. https://proceedings.mlr.press/v97/li19d.html

[28] libexpat. 2025. Expat XML Parser Library. https://libexpat.github.io/.

[29] libpng. 2025. libpng: The PNG Reference Library. http://www.libpng.org/pub/png/libpng.html.

[30] LibTIFF. 2025. LibTIFF: TIFF Library and Utilities. http://www.simplesystems.org/libtiff/.

[31] Zhenhao Luo, Pengfei Wang, Baosheng Wang, Yong Tang, Wei Xie, Xu Zhou, Danjun Liu, and Kai Lu. 2023. VulHawk: Cross-architecture Vulnerability Detection with Entropy-based Binary Code Search. In *30th Annual Network and Distributed System Security Symposium, NDSS 2023, San Diego, California, USA, February 27 - March 3, 2023*. The Internet Society. doi:10.14722/ndss.2023.24415

[32] Luca Massarelli, Giuseppe Antonio Di Luna, Fabio Petroni, Roberto Baldoni, and Leonardo Querzoni. 2019. SAFE: Self-Attentive Function Embeddings for Binary Similarity. In *Detection of Intrusions and Malware, and Vulnerability Assessment - 16th International Conference, DIMVA 2019, Gothenburg, Sweden, June 19-20, 2019, Proceedings (Lecture Notes in Computer Science, Vol. 11543)*. Springer, 309–329. doi:10.1007/978-3-030-22038-9_15

[33] Marius Muench, Jan Stijohann, Frank Kargl, Aurélien Francillon, and Davide Balzarotti. 2018. What You Corrupt Is Not What You Crash: Challenges in Fuzzing Embedded Devices. In *Network and Distributed System Security Symposium (NDSS)*. doi:10.14722/ndss.2018.23166

[34] National Institute of Standards and Technology. 2014. CVE-2014-0160. https://nvd.nist.gov/vuln/detail/cve-2014-0160.

[35] National Institute of Standards and Technology. 2025. National Vulnerability Database (NVD). https://nvd.nist.gov/.

[36] OpenSSL. 2025. OpenSSL: Cryptography and SSL/TLS Toolkit. https://www.openssl.org/.

[37] Kexin Pei, Zhou Xuan, Junfeng Yang, Suman Jana, and Baishakhi Ray. 2020. Trex: Learning Execution Semantics from Micro-Traces for Binary Similarity. *arXiv preprint arXiv:2012.08680* (2020). doi:10.48550/arXiv.2012.08680

[38] Nilo Redini, Aravind Machiry, Ruoyu Wang, Chad Spensky, Andrea Continella, Yan Shoshitaishvili, Christopher Kruegel, and Giovanni Vigna. 2020. Karonte: Detecting Insecure Multi-binary Interactions in Embedded Firmware. In *2020 IEEE Symposium on Security and Privacy (SP)*. 1544–1561. doi:10.1109/SP40000.2020.00036

[39] Liting Ruan, Qizhen Xu, Shunzhi Zhu, Xujing Huang, and Xinyang Lin. 2024. A Survey of Binary Code Similarity Detection Techniques. *Electronics* 13, 9 (2024). doi:10.3390/electronics13091715

[40] Tobias Scharnowski, Nils Bars, Moritz Schloegel, Eric Gustafson, Marius Muench, Giovanni Vigna, Christopher Kruegel, Thorsten Holz, and Ali Abbasi. 2022. Fuzzware: Using Precise MMIO Modeling for Effective Firmware Fuzzing. In *31st USENIX Security Symposium (USENIX Security 22)*. USENIX Association, Boston, MA, 1239–1256. https://www.usenix.org/conference/usenixsecurity22/presentation/scharnowski

[41] Statista Research Department. 2024. *Internet of Things (IoT) connected devices installed base worldwide from 2019 to 2030*. Technical Report. Statista. Available at: https://www.statista.com/statistics/1183457/iot-connected-devices-worldwide/.

[42] Hao Wang, Zeyu Gao, Chao Zhang, Mingyang Sun, Yuchen Zhou, Han Qiu, and Xi Xiao. 2024. CEBin: A Cost-Effective Framework for Large-Scale Binary Code Similarity Detection. In *Proceedings of the 33rd ACM SIGSOFT International Symposium on Software Testing and Analysis (ISSTA 2024)* (Vienna, Austria) *(ISSTA 2024)*. Association for Computing Machinery, New York, NY, USA. doi:10.1145/3650212.3652117

[43] Hongru Wang, Chunfang Li, Lingfei Zhang, and Minyong Shi. 2018. Anti-Crawler strategy and distributed crawler based on Hadoop. In *2018 IEEE 3rd International Conference on Big Data Analysis (ICBDA)*. IEEE, 227–231. doi:10.1109/ICBDA.2018.8367682

[44] Hao Wang, Wenjie Qu, Gilad Katz, Wenyu Zhu, Zeyu Gao, Han Qiu, Jianwei Zhuge, and Chao Zhang. 2022. jTrans: jump-aware transformer for binary code similarity detection. In *Proceedings of the 31st ACM SIGSOFT International Symposium on Software Testing and Analysis* (Virtual, South Korea) *(ISSTA 2022)*. Association for Computing Machinery, New York, NY, USA, 1–13. doi:10.1145/3533767.3534367

[45] Haohuang Wen, Zhiqiang Lin, and Yinqian Zhang. 2020. FirmXRay: Detecting Bluetooth Link Layer Vulnerabilities From Bare-Metal Firmware. In *Proceedings of the 2020 ACM SIGSAC Conference on Computer and Communications Security (CCS)*. ACM, 167–180. doi:10.1145/3372297.3423344

[46] Yuhao Wu, Jinwen Wang, Yujie Wang, Shixuan Zhai, Zihan Li, Yi He, Kun Sun, Qi Li, and Ning Zhang. 2024. Your Firmware Has Arrived: A Study of Firmware Update Vulnerabilities. In *33rd USENIX Security Symposium (USENIX Security 24)*. USENIX Association, Philadelphia, PA, 5627–5644. https://www.usenix.org/conference/usenixsecurity24/presentation/wu-yuhao

[47] Xiaojun Xu, Chang Liu, Qian Feng, Heng Yin, Le Song, and Dawn Song. 2017. Neural Network-Based Graph Embedding for Cross-Platform Binary Code Similarity Detection. In *Proceedings of the 2017 ACM SIGSAC Conference on Computer and Communications Security* (Dallas, Texas, USA) *(CCS '17)*. Association for Computing Machinery, New York, NY, USA, 363–376. doi:10.1145/3133956.3134018

[48] Shouguo Yang, Long Cheng, Yicheng Zeng, Zhe Lang, Hongsong Zhu, and Zhiqiang Shi. 2021. Asteria: Deep Learning-based AST-Encoding for Cross-platform Binary Code Similarity Detection. In *51st Annual IEEE/IFIP International Conference on Dependable Systems and Networks, DSN 2021, Taipei, Taiwan, June 21-24, 2021*. IEEE, 224–236. doi:10.1109/DSN48987.2021.00036

[49] Shouguo Yang, Chaopeng Dong, Yang Xiao, Yiran Cheng, Zhiqiang Shi, Zhi Li, and Limin Sun. 2023. Asteria-Pro: Enhancing Deep Learning-based Binary Code Similarity Detection by Incorporating Domain Knowledge. *ACM Trans. Softw. Eng. Methodol.* 33, 1, Article 1 (Nov. 2023), 40 pages. doi:10.1145/3604611

[50] Jonas Zaddach, Luca Bruno, Aurélien Francillon, and Davide Balzarotti. 2014. AVATAR: A Framework to Support Dynamic Security Analysis of Embedded Systems' Firmwares. In *21st Annual Network and Distributed System Security Symposium, NDSS 2014, San Diego, California, USA, February 23-26, 2014*. The Internet Society. https://doi.org/10.14722/ndss.2014.23229

[51] Binbin Zhao, Shouling Ji, Jiacheng Xu, Yuan Tian, Qiuyang Wei, Qinying Wang, Chenyang Lyu, Xuhong Zhang, Changting Lin, JingZheng Wu, and Raheem Beyah. 2022. A large-scale empirical analysis of the vulnerabilities introduced by third-party components in IoT firmware. In *Proceedings of the 31st ACM SIGSOFT International Symposium on Software Testing and Analysis* (Virtual Event, South Korea) *(ISSTA 2022)*. ACM, 442–454. doi:10.1145/3533767.3534366

[52] Yaowen Zheng, Ali Davanian, Heng Yin, Chengyu Song, Hongsong Zhu, and Limin Sun. 2019. FIRM-AFL: High-Throughput Greybox Fuzzing of IoT Firmware via Augmented Process Emulation. In *28th USENIX Security Symposium (USENIX Security 19)*. Santa Clara, CA, 1099–1114. https://www.usenix.org/conference/usenixsecurity19/presentation/zheng

[53] zlib. 2025. zlib: A Massively Spiffy Yet Delicately Unobtrusive Compression Library. https://zlib.net/.